\documentstyle[fleqn,12pt,epsf]{article}
\begin{document}                  % DO NOT DELETE THIS LINE

       %------------------------------------------------------------------
       % The introductory (header) part of the paper
       %------------------------------------------------------------------

       % The title of the paper. Use \shorttitle to indicate an abbreviated title
       % for use in running heads (you will need to uncomment it).

\title{{\bf Crystallography without crystals I:\\ The common-line method for
assembling a 3D diffraction volume from single-particle scattering}}

       % Authors' names and addresses. Use \cauthor for the main (contact) author.
       % Use \author for all other authors. Use \aff for authors' affiliations.
       % Use lower-case letters in square brackets to link authors to their
       % affiliations; if there is only one affiliation address, remove the [a].

\author{{\bf V. L. Shneerson, A. Ourmazd, and D. K. Saldin}\\
\\
Department of Physics,\\
University of Wisconsin-Milwaukee, P. O. Box 413,\\
Milwaukee, WI 53201, U.S.A.\\
\\
E-mail: valentin@uwm.edu, ourmazd@uwm.edu, dksaldin@uwm.edu}
%\aff[b]{Second affiliation address}

       % Use \shortauthor to indicate an abbreviated author list for use in
       % running heads (you will need to uncomment it).

       % Use \vita if required to give biographical details (for authors of
       % invited review papers only). Uncomment it.

%\vita{Author's biography}

       % Keywords (required for Journal of Synchrotron Radiation only)
       % Use the \keyword macro for each word or phrase, e.g.
       % \keyword{X-ray diffraction}\keyword{muscle}

%\keyword{keyword}

       % PDB and NDB reference codes for structures referenced in the article and
       % deposited with the Protein Data Bank and Nucleic Acids Database (Acta
       % Crystallographica Section D). Repeat for each separate structure e.g
       % \PDBref[dethiobiotin synthetase]{1byi} \NDBref[d(G$_4$CGC$_4$)]{ad0002}

%\PDBref[optional name]{refcode}
%\NDBref[optional name]{refcode}

\maketitle                        % DO NOT DELETE THIS LINE

\begin{abstract}
We demonstrate that a common-line method can assemble a
3D oversampled diffracted intensity distribution suitable for high-resolution
structure solution from a set of measured 2D diffraction patterns, as proposed in 
experiments with an X-ray free electron laser (XFEL) (Neutze {\it et al.}, 2000).
Even for a flat Ewald sphere, we show how the ambiguities due to Friedel's Law may 
be overcome. The method breaks down for photon counts below about 10
per detector pixel, almost 3 orders of magnitude higher than expected for 
scattering by a 500 kDa protein with an XFEL beam focused to a 0.1 $\mu$m 
diameter spot. Even if 10$^3$ orientationally similar diffraction patterns could be 
identified and added to reach the requisite photon count per pixel, the need for
about 10$^{6}$ orientational classes for high-resolution structure determination
suggests that about $\sim$ 10$^9$ diffraction patterns must be recorded. 
Assuming pulse and read-out rates of 100 Hz, such measurements would require
$\sim$ 10$^7$ seconds, i.e. several months of continuous beam time.

\end{abstract}

\newpage
%-------------------------------------------------------------------------
       % The main body of the paper
 
%-------------------------------------------------------------------------
       % Now enter the text of the document in multiple \section's, \subsection's
       % and \subsubsection's as required.

\section{Introduction}

X-ray crystallography is one of the key contributions of the physical sciences
to the life sciences.  Its application to biological, biochemical, and 
pharmaceutical problems continues to enable breakthroughs (Cramer {\it et al.}, 
2001; Gnatt {\it et al.}, 2001) highlighting the importance of structure 
to function.  However, roughly 40$\%$ of biological 
molecules do not crystallize, and many cannot easily be purified. These factors 
severely limit the applicability of X-ray crystallography; although more than 
750,000 proteins have been sequenced, the structures of less than 
10$\%$ have been determined to high resolution (Protein Data Bank, {\it http://www.pdb.org}). 
The ability to determine the structure of individual biological molecules - 
without the need for purification and crystallization - would constitute a 
fundamental breakthrough.  

The confluence of five factors has generated intense interest in single-molecule 
crystallography by short-pulse X-ray scattering: a) The advent of algorithms for 
determining phases from measured diffraction intensities by successive and repeated
application of constraints in real and reciprocal spaces (see e.g. Fienup, 1978; Elser, 2003; 
Millane, 2003), with demonstrations in astronomy (Fienup, 1982); diffractive imaging of 
nanoparticles (Williams {\it et al.}, 2003; Wu {\it et al.}, 2005; Chapman {\it et al.},
2006), biological cells 
(Shapiro {\it et al.}, 2005; Thibault {\it et al.}, 2006); small molecule 
crystallography (Oszl\'{a}nyi and S\"{u}to, (2003); Wu {\it et al.} (2004a); 
surface crystallography (Kumpf {\it et al.}, 2001; Fung {\it et al.}, 2007); and 
protein crystallography (Miao {\it et al.}, 2001; Spence {\it et al.}, 2005); 
b) Development of sophisticated techniques for determining the relative 
orientation of electron microscope images of biological entities, such as cells 
and large macromolecules (see e.g. Frank, 2006);  c) Development of techniques for 
producing beams of hydrated proteins by electrospraying or Raleigh-droplet 
formation (Fenn, 2002; Spence {\it et al.}, 2005);  d) The promise of very 
bright, ultra-short pulses of hard X-rays from X-ray Free Electron Lasers 
(XFELs) under construction in the US, Japan, and Europe (Normille, 2006); e) 
The prospect of overcoming the limits to achievable resolution due to radiation damage
by using short pulses of radiation (Solem and Baldwin, 1982; Neutze {\it et al.},
2000).

It has been suggested (Neutze {\it et al.}, 2000; Hajdu {\it et al.}, 2000;
Abela {\it et al.}, 2007) that an experiment to determine the 
structure of a biological molecule might, in principle, proceed as follows: 
i) A train of individual hydrated proteins is exposed to a synchronized train 
of intense X-ray pulses.  As a single pulse is sufficient to destroy the molecule, 
the pulses (and data collection) must be short compared with the roughly 
50 fs needed for the molecular constituents to fly apart (Neutze {\it et al.}, 
2000; Jurek {\it et al.}, 2004).  ii) The two-dimensional (2D) diffraction patterns 
obtained with single pulses are read 
out, each pattern corresponding to an unknown, random orientation of the molecule.
iii) The relative orientations of the molecule corresponding to 2D 
diffraction patterns (and hence the relative orientations of each diffraction pattern 
in 3D reciprocal space) are determined.  iv) A noise-averaged 3D 
diffracted intensity distribution is constructed.  v) The structure of the 
molecule is determined from the diffracted intensity distribution by an 
iterative ``phasing algorithm'' (Miao {\it et al.}, 2001).

As pointed out by Huldt {\it et al.}, (2003), for this approach 
to succeed in principle, it is necessary to develop a noise-robust algorithm to 
determine the relative orientations of diffraction patterns obtained from 
randomly-oriented individual molecules, to reconstruct the 3D diffracted intensity 
distribution of sufficient quality, and to determine the secondary structure of 
individual biological molecules.

In brief, starting with a collection of noisy 2D diffraction patterns of unknown 
orientation, such a method recovers the 3D electron density of a molecule, providing
a quantitative measure of the reliability of the 
reconstruction. It has been suggested that an algorithm developed for the analogous
problem of the reconstructing a 3D image of a large molecule or nanoparticle from
different projected electron microscope images, the method of common lines, may be 
employed for this task. We investigate the capabilities and limitations of such
an approach for structure recovery from simulated short-wavelength 
diffraction patterns of a small (10-residue) synthetic protein, Chignolin 
(Protein Data Bank Entry 1UAO).

Starting with 630 simulated, noise-free 2D diffraction patterns of 0.1 \AA \ wavelength 
X-rays from random orientations of the molecule, we show that such an algorithm is 
able to recover the electron density distribution of the (small) test protein molecule, 
Chignolin, up to about 1 \AA \ resolution with a fidelity measured by a correlation 
coefficient of 0.7 between the model and recovered electron 
density distributions. This constitutes the first demonstration of an integrated
algorithm able to perform all the tasks necessary to extract a molecular
electron density from a set of 2D diffraction patterns of 
random unknown orientations. We have also investigated the limits of the algorithm
with respect to shot noise (modeled by Poisson statistics) in the detected signal. 
Our results show that the common-line method requires at least 10 photons/pixel.
This is at least two orders of magnitude higher than the anticipated signal levels
from the LCLS XFEL currently under construction.

The algorithm consists of three primary modules: a) Determination of the relative 
orientations of the measured diffraction patterns in 3D reciprocal space; b) From the
resulting irregular distribution of diffracted intensities, application of a gridding 
algorithm to generate data on a uniform rectilinear grid in reciprocal space; c) 
Application to this gridded data of a 3D iterative algorithm to find the phases 
associated 
with the grid intensities, and recover the 3D electron density of the molecule. For
an experiment which provided independent information about the orientations of
the sample, steps b) and c) were previously implemented by Chapman {\it et al.} (2006).  

\section{Determination of the Relative Orientations of the Diffraction Patterns}

In the following, we assume that the X-ray energy is high enough, and the solid 
angle subtended by the diffraction pattern at the sample small enough that it 
is a reasonable approximation to consider each diffraction pattern as a planar 
central section through the 3D reciprocal space of the molecule. In practice, 
this is valid for X-ray wavelengths of about 
0.1 \AA. Then, the problem reduces to determining the 
relative orientations of these planar sections from the data in the diffraction patterns 
alone without any knowledge of the structure of the molecule. For longer wavelengths, as 
pointed out previously (e.g. Huldt {\it et al.}, 2003, Chapman, 2007), it will 
be necessary to take account of the curvature of the Ewald sphere, when the 
common lines become arcs of a circle rather than straight lines. The extra complexity of 
identifying such arcs may be offset by two factors: (1) the avoidance of ambiguities
stemming from the duplication of intensities in the same 2D diffraction pattern, due to
Friedel's Law (see below); and (2) the possibility of 
determining all three Euler angles relating the orientations of any two
diffraction patterns from just their mutual common line (Huldt, {\it et al.}, 2003), 
if the extra parameter of the radius of curvature of the common arcs may be determined 
with sufficient accuracy.

Our approach is inspired by the analogous problem of reconstructing the 3D structure
of a macromolecule or nanoparticle from electron microscope images representing
projections of copies of the object along random directions (Frank, 2006). This problem 
has been solved by exploiting the central section
theorem (Farrow and Ottensmeyer, 1992), and has been developed most notably by these
authors and also previously by van Heel (1987); and Goncharov et al., (1987). From the 
{\it projection-slice theorem}, the Fourier transform of a 2D (i.e. projected) image 
is a central slice through the complex 3D reciprocal space of the 3D object. 
Any two central sections intersect along a line. This allows partial alignment of
the two central sections with respect to each other. Specifically, determination 
of the gradients of this common line relative to, say, 2D Cartesian coordinate 
systems in the planes of each of the central sections allows two of the three
Euler angles specifying the relative orientations of these central sections to be
deduced. Using this procedure, it 
is generally possible to determine six of the nine interplanar Euler angles 
between {\it three} independent diffraction patterns. Knowledge of these six 
Euler angles allows the remaining three to be deduced by geometrical 
construction.  

We point to one important difference between an application of the common-line approach 
to images (as in 3D electron microscopy) and diffraction patterns. The images constitute
projections of the object in real space. Some applications of the central section 
theorem have been performed in reciprocal space, exploiting the fact that 2D Fourier 
transforms of these images
yield moduli and phases of {\it complex} amplitudes on central sections through 
reciprocal space. Sinograms of the data of any two diffraction patterns allow 
the unique identification of a pair of Euler angles relating the two central sections 
in 3D reciprocal space (Frank, 2006).
In contrast, in our problem, the raw experimental data are diffracted intensities, 
and direct information is available only about the {\it moduli} of the complex 
amplitudes in reciprocal space. 

Friedel's Law of crystallography suggests that the intensity distribution along a 
radial line through the center of each diffraction pattern is the same as one rotated 
relative to it by 180$^{o}$. This means that, for a flat Ewald sphere, the determination 
of the pair of Euler angles from common lines is uncertain by $\pm$ 180$^{o}$. 
Any significant curvature of the Ewald sphere removes this
ambiguity, but even for a flat Ewald sphere, this uncertainty may be resolved through 
consistency conditions amongst the Euler angles, as shown below. 

\subsection{Determination of two Euler angles between two intersecting 
diffraction patterns}

\begin{figure}
\begin{center}
\epsfxsize=2.9in
\epsffile{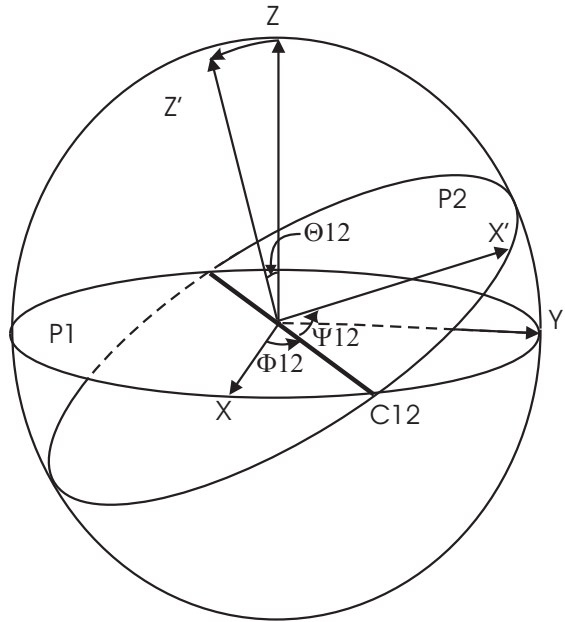}
\end{center}
\caption{{\it Transformation of central section $P1$ into $P2$, by rotation
through Euler angles $\Phi$12, $\Theta$12, and $\Psi$12}.} 
\label{euler}
\end{figure}

Fig. \ref{euler} illustrates the reciprocal-space geometry of two central sections, 
representing diffraction patterns, P1 and P2. Let the Euler angles relating 
P1 and P2 be $\Phi12$, $\Theta12$, and $\Psi12$. Consider three Cartesian axes X, Y, 
and Z, where X and Y lie in the plane of P1, and Z is normal to it. Diffraction pattern 
P2 is related to P1 by a set of three rotations. 
The initial rotation is through the azimuthal angle $\Phi12$ about the Z axis.
Next follows a rotation through $\Theta12$ about the X axis obtained after 
the first rotation. Let us denote this axis by C12. The final rotation is 
through $\Psi12$ about the new Z axis, denoted Z'. It is clear from 
the figure that C12 is the line of intersection between P1 and P2, i.e
the common line.
   
\begin{figure}
\begin{center}
\epsfxsize=2.9in
\epsffile{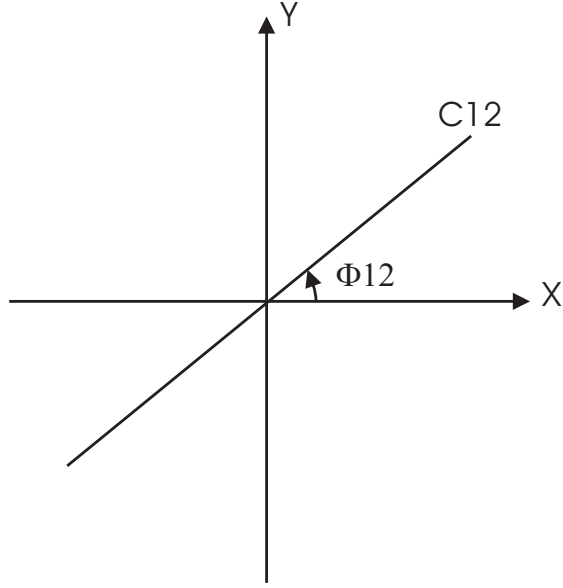}
\end{center}
\caption{{\it Orientation of common line $C12$ relative to the Cartesian
axes ($X$,$Y$) in central section $P1$.}}
\label{p1}
\end{figure}

The orientation of the common line C12 relative to the (X,Y) Cartesian axes in the 
plane of P1 is shown in Fig. \ref{p1}. The gradient m12 of C12 with respect to 
the (X,Y) axes in the plane of P1 is given by
\begin{eqnarray}
m_{12}=\tan{(\Phi12)} 
\end{eqnarray}
and hence the Euler angle
\begin{equation}
\Phi12 = \arctan{(m_{12})}
\end{equation}
can be determined if the common line C12, and hence its gradient in the plane of 
P1 relative to the Cartesian axes (X,Y) can be identified.

\begin{figure}
\begin{center}
\epsfxsize=2.9in
\epsffile{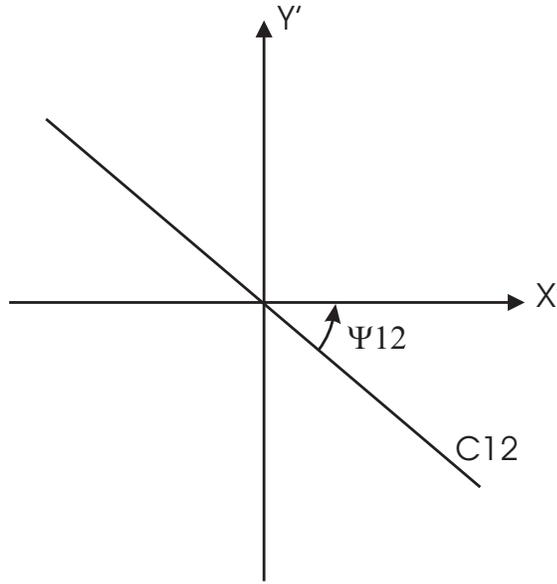}
\end{center}
\caption{{\it Orientation of common line $C12$ relative to the Cartesian
axes ($X'$,$Y'$) after the Euler-angle rotations in the diffraction pattern$P2$.}}
\label{p2}
\end{figure}

Now note that since C12 is the common line, it must also be contained in the 
plane of P2 after the Euler-angle rotations. Its orientation relative to the 
Cartesian axes (X',Y') in the plane of P2 is depicted in Fig. \ref{p2}. Its gradient 
m21 relative to the axes (X',Y') in P2 is
\begin{eqnarray}
m_{21}=-\tan{(\Psi12)} 
\end{eqnarray}
and hence the Euler angle
\begin{equation}
\Psi12 = \arctan{(-m_{21})}
\end{equation}
may be determined if the common line C12, and hence its gradient in the plane of P2 
relative to the Cartesian axes (X',Y') can be identified in the diffraction 
pattern in that plane.

Given two diffraction patterns, a pairwise
numerical comparison ({\it sinogram} comparison, Frank, 2006) of the intensity 
distributions along radial directions 
of the two patterns may be conducted. An automated criterion,
such as an R-factor, monitors the degree of agreement. An exhaustive search is 
performed of all pairs of radial distributions of the intensities on the two
patterns. A global minimum of the R-factor is assumed to determine the
common line. For the diffraction patterns P1 and P2 above, denote the common line
by C12. This gives estimates of the Euler angles $\Phi$12 and $\Psi$12.

\begin{figure}
\epsfxsize=3.0in
\epsffile{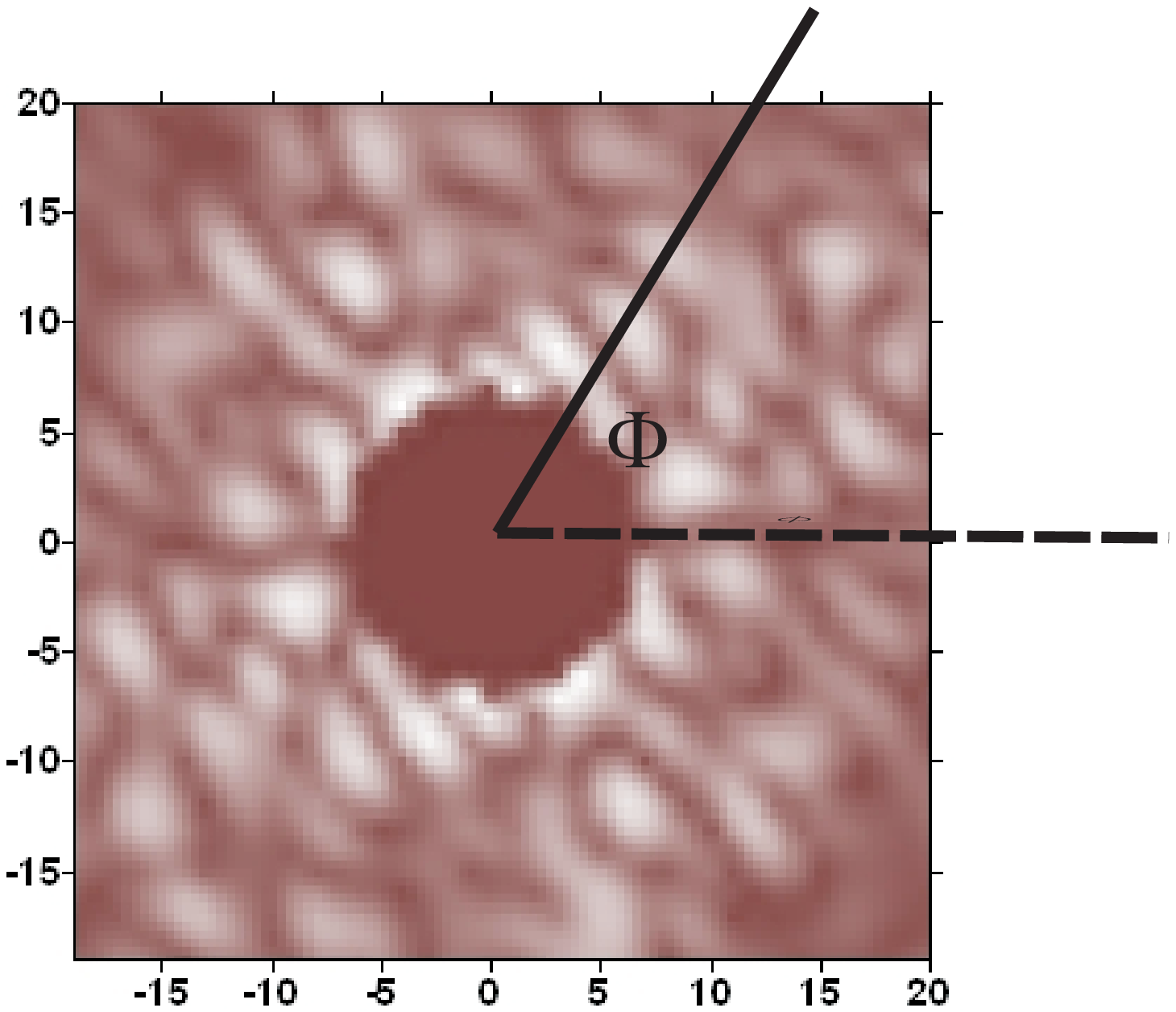}
\epsfxsize=2.7in
\epsffile{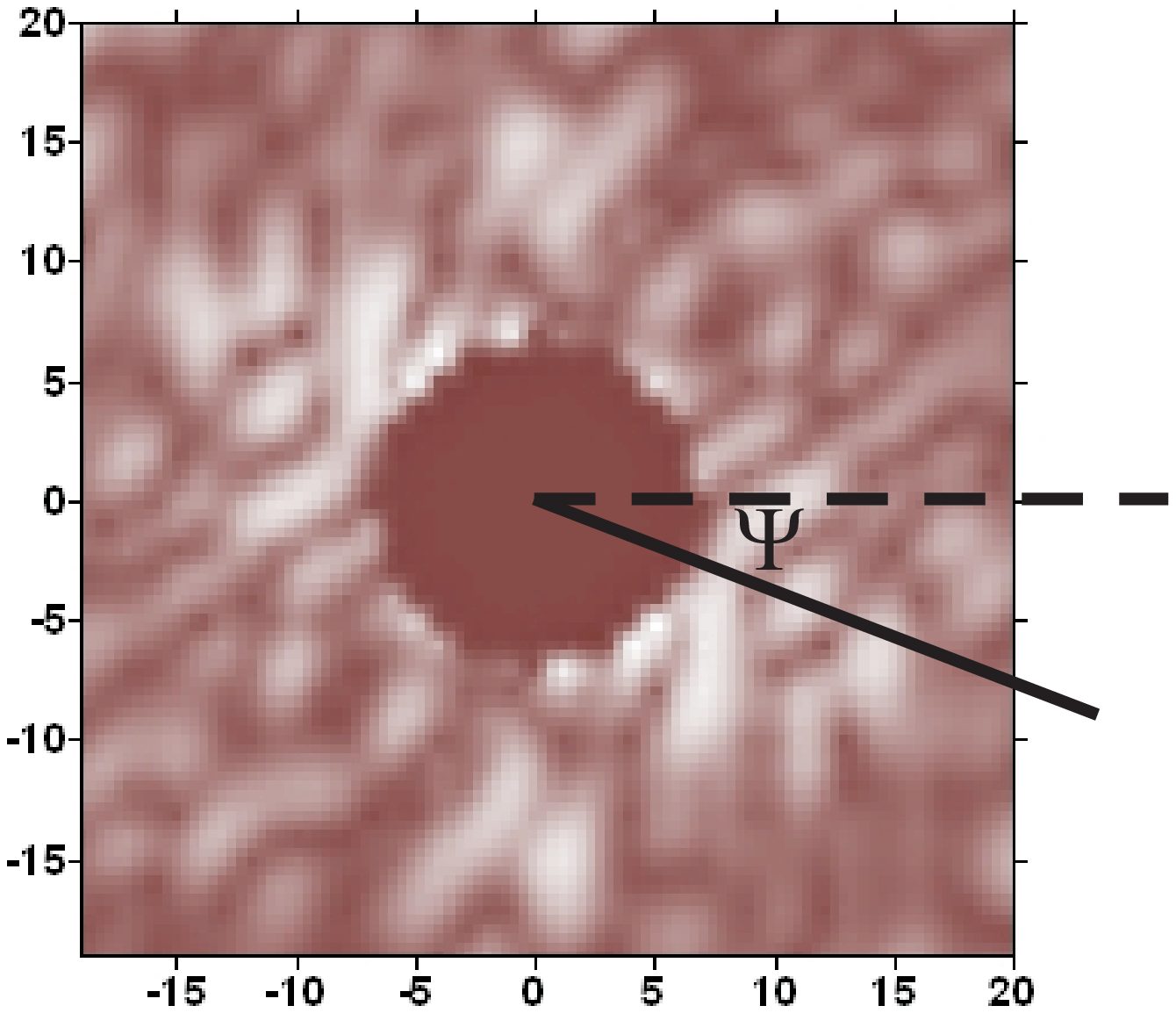}
\caption{{\it Identification of the common line in two typical simulated diffraction 
patterns from a model of the protein Chignolin, leading to a determination of
the azimuthal Euler angles $\Phi$ and $\Psi$ relating the 3D orientations of the
diffraction patterns.}}
\label{dps}
\end{figure}

\begin{figure}
\begin{center}
\epsfxsize=3.4in
\epsffile{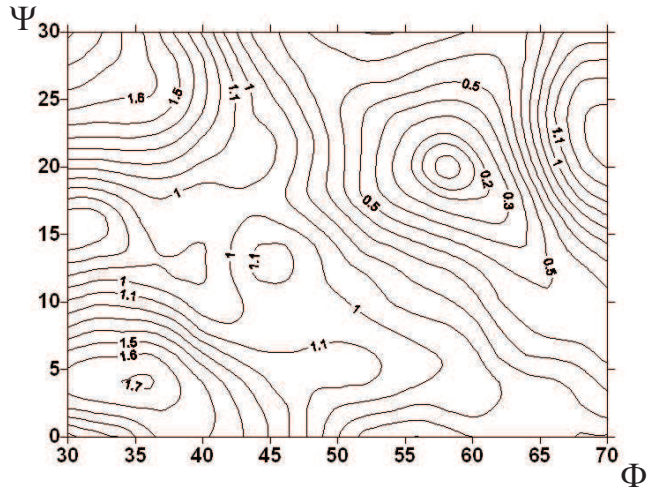}
\end{center}
\caption{{\it Contour map of sinogram comparisons between the two diffraction
patterns of Fig. \ref{dps} in the vicinity of the global minimum at 
$\Phi$=58$^{o}$, $\Psi$=21$^{o}$.}
}
\label{contour}
\end{figure}

Fig. \ref{dps}(a) and (b) show two simulated diffraction patterns (40$\times$40 pixels)
from random orientations of our test protein, Chignolin. The maximum lateral
wavevector in the direction of the x-axis was 10 times the Nyquist frequency
for the assumed lateral extension of the protein (16 \AA). This corresponds
to a reciprocal-space length of $q=2\pi(10)/16=3.93$ \AA$^{-1}$. The wavevector, $k$,
of 124 keV hard X-rays is about 63 \AA$^{-1}$. The scattering angle 
corresponding to the middle of an edge of the square diffraction pattern 
was calculated from $2\arcsin(q/2k)=3.2^{o}$. The central part of each diffraction 
pattern contains high intensities of relatively low detail, but several 
orders of magnitude stronger than in the outer parts of the pattern containing 
the high-resolution structural information. A numerical search for 
the common lines between the two patterns of Fig. \ref{dps} was performed by pairwise
comparisons of the radial intensity distributions from the two patterns in angular
steps of 1$^{o}$, excluding the pixels within a central high-intensity disc of 
7-pixel radius corresponding to a scattering angle of $\sim$ 1$^{o}$. Effectively, 
the values of the azimuthal Euler angles 
$\Phi$12 and $\Psi$12 were identified by a contour plot of the form shown 
in Fig. \ref{contour} The identified common lines are also shown in Fig. 
\ref{dps}(a) and (b).
Due to the Friedel Law degeneracy mentioned above, any 180$^{o}$ range of 
azimuthal angles would be expected to contain such a minimum. For convenience,
we perform numerical searches for $\Phi$12 and $\Psi$12 angles over an azimuthal 
angle range of 0 to 180$^{o}$. Then Friedel's Law suggests equally valid
values for these angles of $\Phi$12 + 180$^{o}$ and $\Psi$12 + 180$^{o}$, 
respectively. Without phase information, it is impossible to tell from the diffraction
data alone, which 
of the two values of each angle is ``correct''. 

In the case of the flat Ewald sphere considered here, it is not 
possible to determine the Euler angle $\Theta$12 between the normals to these 
planes with the data in the diffraction patterns $P1$ and $P2$ alone. In order to 
determine that angle, it is necessary to have diffraction data in at least one more 
distinct reciprocal-space plane, which intersects the planes of $P1$ 
and $P2$ along two further distinct common lines. 

\subsection{Determination of all nine Euler angles relating three general
central sections}

\begin{figure}
\begin{center}
\epsfxsize=2.9in
\epsffile{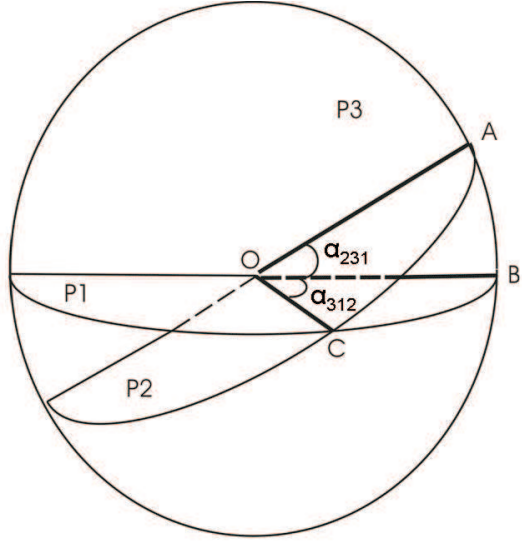}
\end{center}
\caption{{\it Geometrical construction for determining the relative Euler
angle $\Theta12$ between diffraction patterns $P1$ and $P2$, given the six Euler azimuthal 
angles $\Phi$ and $\Psi$ relating $P1$, $P2$ and $P3$.}} 
\label{sph_trig}
\end{figure}

Let $P3$ denote a third diffraction pattern (Fig. \ref{sph_trig}). Since 
each diffraction pattern forms a central section through reciprocal space, each pair of 
diffraction patterns intersect along a common line, with the three common lines 
intersecting at the origin ($O$ in Fig. \ref{sph_trig}). Denote the Euler angles 
specifying the transformation of the plane of $P2$ to that of $P3$ by 
($\Phi23$,$\Theta23$,$\Psi23$), and those transforming plane of $P3$ to that of $P1$ by   
($\Phi31$,$\Theta31$,$\Psi31$). In the notation of Fig. \ref{sph_trig},
the common line between $P1$ and $P2$ is denoted by $OC$, that
between $P2$ and $P3$ by $OA$ and that between $P3$ and $P1$ by $OB$, with 
$A$, $B$, and $C$ representing points on the surface of a unit sphere 
centered on $O$.

By analogy with the method described in the last section, a comparison of 
the diffraction intensities of $P2$ and $P3$ can determine
the Euler angles $\Phi23$ and $\Psi23$. Likewise comparison of the
data of $P3$ and $P1$ can determine the angles $\Phi31$ and
$\Psi31$. This leaves only three angles to be determined: $\Theta12$ between 
$P1$ and $P2$; $\Theta23$ between $P2$ and $P3$; and $\Theta31$,
between $P3$ and $P1$.

The geometrical construction of Fig. \ref{sph_trig} shows that the
remaining Euler angles are the vertex angles $\angle ACB$, 
$\angle BAC$, and $\angle CBA$ of
the spherical triangle $ABC$ on the surface of the unit sphere.
Also note that the lengths of the sides of this spherical triangle
(the arcs $CB$, $BA$, and $AC$) are equal to the sums of angles 
$\Psi31+\Phi12=\alpha_{312}$ (say), $\Psi12+\Phi23=\alpha_{123}$ 
(say), and $\Psi23+\Phi31=\alpha_{231}$ (say), respectively (expressed
in radians). (For example, if we consider a transformation from plane 3 to 
plane 1 followed by one from plane 1 to plane 2, then the third Euler angle
in the former transformation ($\Psi31$) and the first Euler angle in the
latter transformation ($\Phi12$) involve rotations in the same plane, that of $P1$.) 

The {\it cosine rule} of spherical trigonometry gives
\begin{equation}
\cos{(AB)}=\cos{(CA)}\cos{(CB)} + \sin{(CA)}\sin{(CB)} \cos{(\angle ACB)}
\end{equation}
that is, 
\begin{eqnarray}
\cos{\alpha_{231}}=\cos{\alpha_{123}} \cos{\alpha_{312}}
+ \sin{\alpha_{123}} \sin{\alpha_{312}} \cos{(\Theta12)}
\end{eqnarray}
and thus 
\begin{equation}
\Theta12=\arccos{\left[\frac{\cos{\alpha_{231}}-\cos{\alpha_{123}} \cos{\alpha_{312}}}
{\sin{\alpha_{123}}\sin{\alpha_{312}}}\right]}.
\end{equation}
This expression was obtained by Goncharov {\it et al.} (1987) by a different argument.

Generalizing this result for a triplet of diffraction patterns $i$, $j$, and $k$, the angle 
$\Theta_{ij}$ between $i$ and $j$, is given by
\begin{equation}
\Theta_{ij}=\arccos{\left[\frac{\cos{\alpha_{jki}}-\cos{\alpha_{ijk}}
\cos{\alpha_{kij}}}
{\sin{\alpha_{ijk}}\sin{\alpha_{kij}}}\right]}
\label{Theta_ij},
\end{equation}
where
\begin{equation}
\alpha_{ijk}=\Psi_{ij}+\Phi_{jk},
\label{alpha_ijk}
\end{equation}
\begin{equation}
\alpha_{kij}=\Psi_{ki}+\Phi_{ij},
\label{alpha_kij}
\end{equation}
and
\begin{equation}
\alpha_{jki}=\Psi_{jk}+\Phi_{ki}
\label{alpha_jki}
\end{equation}
with $k$ a third plane.

The above analysis shows that, provided the Euler angles $\Phi$ and $\Psi$ specifying the 
directions of common lines between any three sets of diffraction patterns $i$, 
$j$, and $k$ are determined (e.g., by comparisons of sinograms from the diffraction 
patterns), the Euler angles $\Theta$ about the ``hinge axes'' formed by the common 
lines amongst those diffraction patterns can be deduced by the analytic formula 
(\ref{Theta_ij}).

\subsection{Removal of the ambiguities due to Friedel's Law}

As pointed out in the previous section, the Euler angles $\Phi$ and $\Psi$ may be 
determined from the diffraction pattern data only to modulo 180$^{o}$, due to Friedel's Law. 
Thus, in addition to initial values (in the range 0 to 180$^{o}$) assigned to these 
angles by the automated numerical sinogram comparison, one must also consider as 
possible values of these angles, $\Phi$ + 180$^{o}$ and $\Psi$ + 180$^{o}$, 
respectively. The possibility of two values for each of the three $\Phi$ angles and 
two for each of the three $\Psi$ angles, implies four possible values of each of 
the $\alpha$ angles in expressions (\ref{alpha_ijk}), (\ref{alpha_kij}), and 
(\ref{alpha_jki}). Since three distinct $\alpha$ angles enter into the formula 
(\ref{Theta_ij}), there are 4$^{3}$=64 possible values of $\Theta_{ij}$ 
for a given set of $\Phi$ and $\Psi$ angles deduced from three different diffraction 
patterns. 

In fact, this is not the case. Many combinations of $\Phi$ and $\Psi$ give rise to the 
same $\Theta_{ij}$, and a large number of combinations result in arguments of the 
{\it arccos} function in (\ref{Theta_ij}) outside the range of -1 to 1, giving no 
geometrically meaningful solution at all. This eliminates all but two sets of 
the three $\Theta$ angles. The remaining 
ambiguity is due to the well known enantiometric ambiguity of molecular structures
that give rise to the same diffraction intensities. This ambiguity is
impossible to resolve from the diffraction data alone. An arbitrary but consistent choice 
of one of the two sets of $\Theta$ angles produces one of the
two enantiomers of the structure.

A concrete example from a simulation of three diffraction patterns, 1, 2, and 3
from random orientations of the same molecule is illustrative. The 
$\Phi$ and $\Psi$ angles of Table 1 were determined by
numerical comparisons of sinograms of the three patterns. 

\begin{table}
\begin{center}
\begin{tabular}{||l|c|c||}
\cline{1-3}
Pair & $\Phi$ & $\Psi$ \cr
\cline{1-3}
(1,2) & 16.0 & 108.0 \cr
\cline{1-3}
(2.3) & 173.0 & 174.0 \cr
\cline{1-3}
(3,1) & 121.0 & 132.0 \cr
\cline{1-3}
\end{tabular}
\end{center}
\caption{$\Phi$ and $\Psi$ Euler angles (in degrees) relating diffraction
patterns from three random orientations of the molecule, as determined by
numerical sinogram comparisons.}
\label{table1}
\end{table}

Substituting all 64 combinations of $\Phi$ and $\Phi + \pi$, and 
$\Psi$ and $\Psi + \pi$ into
Eqs.(\ref{Theta_ij})-(\ref{alpha_jki}) results in 46 combinations
with values for the cosine of the relevant angle $\Theta$ lying 
outside the range -1 to 1. Nine of the 64 combinations give rise to the 
$\Theta$ angles in Table 2.

\begin{table}
\begin{center}
\begin{tabular}{||l|c||}
\cline{1-2}
Pair & $\Theta$ \cr
\cline{1-2}
(1,2) & 114.2\cr
\cline{1-2}
(2,3) & 144.4\cr
\cline{1-2}
(3,1) & 103.4\cr
\cline{1-2}
\end{tabular}
\end{center}
\caption{One set of three (hinge) angles $\Theta$ (in degrees) between the three 
diffraction patterns oriented in 3D reciprocal space, deduced from Eq.
(\ref{Theta_ij}).}
\label{table2}
\end{table}

Another 9 combinations give rise to the values for the $\Theta$ angles in
Table 3.

\begin{table}
\begin{center}
\begin{tabular}{||l|c||}
\cline{1-2}
Pair & $\Theta$ \cr
\cline{1-2}
(1,2) & 19.1\cr
\cline{1-2}
(2,3) & 12.1\cr
\cline{1-2}
(3,1) & 159.5\cr
\cline{1-2}
\end{tabular}
\end{center}
\caption{Another set of values of the same $\Theta$ angles (in degrees) as in Table
\ref{table2}, as determined by the same method. This solution corresponds to the
enantiomer structure.}
\label{table3}
\end{table}

It turns out that the two sets of values of the $\Theta$ angles determined 
by this method correspond to the two enantiometric solutions referred to 
above. Thus the method described rules out Friedel pair combinations
of common-line directions that are unphysical, producing just the
two enantiomers consistent with the diffraction data.

\subsection{Averaging and Self-Consistency Checks}
  
A particular angle $\Theta_{ij}$ may be estimated from (\ref{Theta_ij}) by taking as the 
third diffraction pattern $k$ any one of the $N-2$ other diffraction patterns. Each choice 
of third diffraction pattern will yield two possible (usually widely separated) values of 
$\Theta_{ij}$, corresponding to the two possible enantiomers. This time, since 
we have already chosen an enantiomer in our previous estimate of $\Theta_{ij}$ using a 
different third diffraction pattern, we choose the solution that is closest to the previously
selected value of $\Theta_{ij}$, i.e., the same enantiomer. The finally assigned
value of this angle will be the average of these values computed by (\ref{Theta_ij}) 
via all possible third planes $k$, namely:
\begin{equation}
\Theta_{ij}=\frac{1}{N-2} \sum_{k\neq i,k\neq j} 
\arccos{\left[\frac{\cos{\alpha_{jki}}-\cos{\alpha_{kij}}\cos{\alpha_{ijk}}}
{\sin{\alpha_{kij}}\sin{\alpha_{ijk}}}\right]}.
\label{NTheta}
\end{equation}

The calculations of the $\Theta$ angles via
(\ref{NTheta}), including 
the tests of enantiomeric consistency, are very rapid. The bulk of the computational
time involves the sinogram comparisons for diffraction pattern pairs. The time
for these computations scales as the total number of pairs amongst $N$ diffraction 
patterns, namely, $N(N-1)/2$. To save computational time for the 630 diffraction 
patterns, we divided them into sets of about 10 diffraction patterns each. 
So long as two diffraction patterns are common to each of these sets of about 10, 
the method determines the relative orientations of all diffraction 
patterns relative to these two for a given enantiomer, with a computational time 
saving of a factor of approximately (630/10)$^{2} \simeq$ 4000.  

A method of sinogram matching determines common-line directions by comparison
between pairs of projections/diffraction patterns at a time. Farrow and 
Ottensmeyer (1992) have suggested a method of simultaneously taking account of 
data from all available projections by means of quarternion mathematics. 
We propose here an alternative method of ensuring that all determined Euler angles 
are consistent with the data of all available diffraction patterns. With 
noisy data, such a self-consistency condition may even help reduce some of the errors 
due to noise. Consider any three noncoplanar diffraction
patterns, $i$, $j$, and $k$. Then
\begin{equation}
R(\Phi_{ij}, \Theta_{ij}, \Psi_{ij})
R(\Phi_{jk}, \Theta_{jk}, \Psi_{jk})
R(\Phi_{ki}, \Theta_{ki},  \Psi_{ki}) = I,
\label{r1}
\end{equation}
where $R$ is the 3D rotation matrix, which transforms plane $i$ to
plane $j$ in 3D reciprocal space, and $I$ is the 3D unit matrix. 
Since 
$R(\Phi_{jk}, \Theta_{jk}, \Psi_{jk})^{-1}
=R(\Phi_{kj}, \Theta_{kj}, \Psi_{kj})$, and
$R(\Phi_{ki}, \Theta_{ki}, \Psi_{ki})^{-1}
=R(\Phi_{ik}, \Theta_{ik}, \Psi_{ik})$, Eq.(\ref{r1}) may be rewritten
\begin{equation}
R(\Phi_{ij}, \Theta_{ij}, \Psi_{ij}) =
R(\Phi_{ik}, \Theta_{ik}, \Psi_{ik})
R(\Phi_{kj}, \Theta_{kj}, \Psi_{kj}), \ \ \ \ \ \ \ \ \forall k. 
\label{r2}
\end{equation}
If the $\Phi$ and $\Psi$ angles on the RHS of (\ref{r2}) have been found 
by sinogram matching, and the $\Theta$ angles on the RHS via 
Eq.(\ref{NTheta}), Eq. (\ref{r2}) may be used to update the Euler
angles $\Phi_{ij}$, $\Theta_{ij}$, and $\Psi_{ij}$ on its LHS. 
Since, for given planes $i$ and $j$, there are $N-2$ other planes $k$,
these angles may be calculated independently from $N-2$ equations of the form 
(\ref{r2}), and the values averaged. A different pair of planes $ij$ can then be 
selected and the procedure repeated to update the Euler angles on the LHSs
of (\ref{r2}) relating all pairs of planes.

To summarize this section, we have described a detailed procedure for orientating
in 3D reciprocal space, a large number of diffraction 
patterns from random unknown orientations of an object {\it without any knowledge of the 
structure of the object}. The recovery of
a molecular electron density from such data requires the determination
of the {\it phases} associated with these intensities. This may be done
by the method of {\it oversampling} (Miao {\it et al.}, 2001), involving
iterative Fourier transformations of the data from reciprocal to
real space, and applications of appropriate constraints in each of the
spaces. A conventional fast Fourier transform (FFT) algorithm (Cooley and Tukey, 1965) 
requires data on a regular Cartesian grid in each space. Thus, it is
necessary to perform a {\it gridding} operation in 3D reciprocal space
to prepare the data for such an iterative {\it phasing} algorithm.

\section{Forming a regular 3D diffracted intensity grid from randomly inclined 
central sections}

We perform this 3D gridding operation by means of the MATLAB routine, 
{\it griddata3}. This routine fits a hypersurface of the form $w=f(x,y,z)$ to the 
irregularly spaced data from the randomly inclined central sections in reciprocal
space, using a tessellation-based linear interpolation, which incorporates the 
method of Delaunay triangulation (Delaunay, 1934). The density of the uniform 3D 
grid points 
was chosen to ensure {\it oversampling} with respect to the Nyquist criterion for 
an object of the size of our test molecule. For the purposes of our present
simulation, where the small test protein Chignolin is known to be smaller than
a cube of linear dimension 16 \AA, we take a reciprocal space sampling 
corresponding
to the Nyquist frequency of a cube of double this linear dimension, namely 32 \AA. 
That is, sampling frequency of the uniform rectilinear 3D reciprocal space grid
is twice the Nyquist frequency corresponding to the diameter of the object
in each of the three linear dimensions. 

\begin{figure}
\begin{center}
\epsfxsize=2.9in
\epsffile{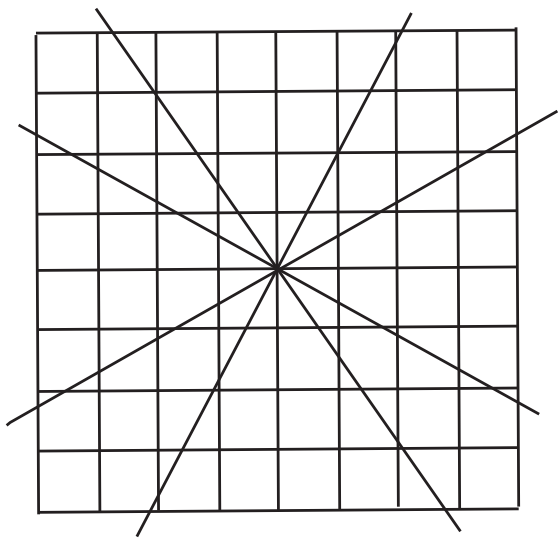}
\end{center}
\caption{{\it 2D representation of the 3D gridding process. Data from regular grids 
on randomly oriented central sections are interpolated onto a regular rectilinear 
3D grid convenient for a fast Fourier transform routine.}}
\label{gridding}
\end{figure}

As the test was performed on simulated 
data, the efficiency of the determination of the relative orientations of the 
simulated diffraction patterns and of the gridding algorithm could be evaluated by 
comparing the diffraction data on the final uniform 3D Cartesian grid with 
diffraction 
intensities calculated directly on the same grid from the PDB atomic data.
The usual X-ray R-factor was used to compare the two datasets. For
our simulation of 630 diffraction patterns from the protein Chignolin,
we obtained an R-factor value of 0.04, indicating a high fidelity for the 
the orientation and gridding process.    

\section{Phasing of the diffraction data and the recovery of the 3D
molecular electron density}

The determination of the phases associated with the gridded diffraction data, and 
hence the 3D molecular electron density was performed by a combination of an 
iterative {\it oversampling} algorithm (Miao {\it et al.}, 2001), which
successively imposes constraints/modifications of the electron density in real 
space through {\it object domain operations} (ODO) (Fienup, 1978; Oszl\'{a}nyi and 
S\"{u}to, 2003) and in reciprocal space (Oszl\'{a}nyi and S\"{u}to, 2004).

The 3D Fourier transform of the gridded diffraction
{\it intensities} yields the 3D autocorrelation function of
the molecular electron density. Since the extent of the autocorrelation map 
is twice that of the electron density map, the approximate spatial extent    
of the molecular electron density can be found directly from the
diffraction intensities (Marchesini {\it et al.}, 2003). 

\begin{figure}
\begin{center}
\epsfxsize=5.5in
\epsffile{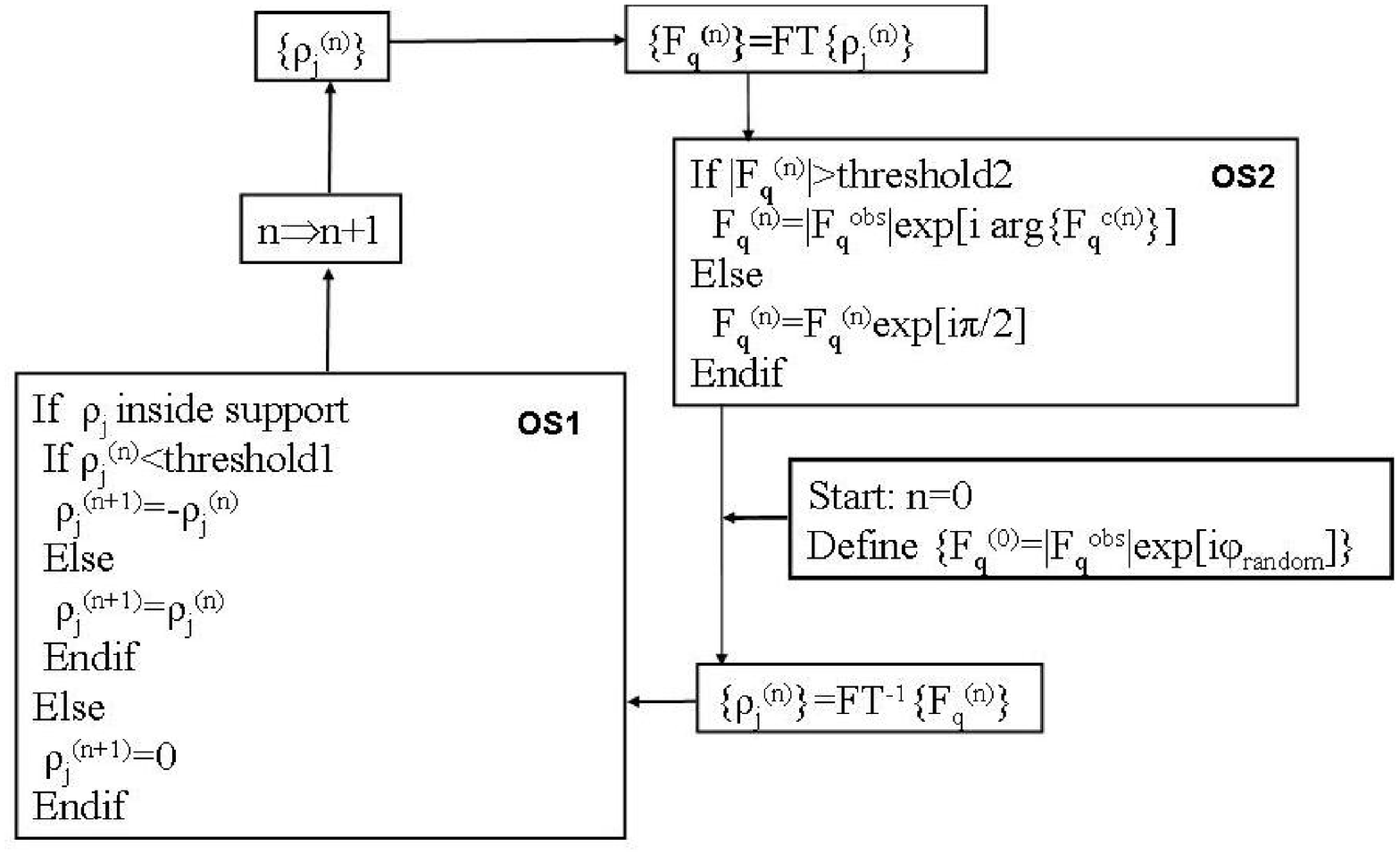}
\end{center}
\caption{{\it Flow chart and pseudo code of the iterative phasing algorithm}}
\label{flow_os}
\end{figure}

A flow chart and pseudo code of our iterative phasing algorithm is shown in
Fig. \ref{flow_os}. The square roots of the gridded diffraction intensities are
assumed proportional to the protein structure factors $F_{\bf q}$, say, where
a reciprocal-space vector ${\bf q}$ is defined by
\begin{equation}
{\bf q} = h {\bf b}_{1} + k {\bf b}_{2} + l {\bf b}_{3}
\end{equation}
where the unit vectors ${\bf b}_{i}$ ($i=1,2,3$) of the reciprocal space are
defined by the usual relationships
\begin{equation}
{\bf b}_{i} \cdot {\bf a}_{j} = \delta_{ij}
\end{equation}
with respect to real-space unit vectors ${\bf a}_{j}$ so chosen as to define a
3D volume expected to contain the molecule. Since the phases associated
with these structure factors are initially unknown, we begin by assigning
random phases to those structure factors $F_{\bf q}$ corresponding to values
of the Laue index $l \ge 0$. Assumption of Friedel's Law,
\begin{equation}
F_{-{\bf q}} = F^{*}_{\bf q}
\end{equation}
then allows the assignment of complex structure factors for $l < 0$. An (inverse) 
FFT algorithm calculates an initial 3D electron density distribution, whose
reality (in the mathematical sense) is assured by the above Friedel relationship
amongst the structure factors. In general, the computed electron density is spread over 
a real-space volume larger than that of the molecule.

A support constraint is now applied in real space by setting to zero the electron
density outside the volume expected to be occupied by the protein (Fienup, 1978).
In addition, the electron density within the expected volume of the protein is
modified according to the {\it charge flipping} prescription of Oszl\'{a}nyi and 
S\"{u}to (2003) (which was shown by Wu {\it et al.} (2004b) to be a special case
of Fienup's (1982) output-output algorithm with feedback parameter $\beta = 2$). According 
to the charge flipping prescription, electron density values that exceed a 
certain threshold $\delta$ are unmodified, while the signs of those below this threshold 
are reversed. The value of this threshold is chosen to optimize the progress
of the algorithm, as monitored by an R-factor between the gridded ``experimental''
structure factors and those calculated from a Fourier transform of the electron
density recovered by the algorithm. (The value for $\delta$ taken in practice 
was typically around 10\% of the maximum electron density.) A Fourier transform of the
modified electron density specifies the same distribution in reciprocal space.
The continued reality of this modified electron density ensures the resulting
calculated structure factors have phases satisfying Friedel's Law.

A different threshold is employed to divide the reciprocal-space amplitudes
into {\it strong} and {\it weak} reflections. The magnitude of the threshold 
amplitude was again monitored by the same R-factor as for the real-space threshold
above. The optimum division was found when 55\% of the weakest reflections were
classified as weak. A reciprocal-space constraint is applied to the strong 
reflections: their amplitudes (or moduli) are replaced by the square roots of 
the corresponding {\it measured} intensities, while 
retaining the phases from the Fourier transform operation. As for the weak
reflections, their moduli are left unchanged, but their phases are shifted by
$\pi/2$. The resulting set of complex structure factors is then 
subject to an inverse Fourier transformation, which yields another
real-space electron distribution. This is modified in the same way as before,
and the whole process repeated for several iterations.

This algorithm constrains the solution to be consistent with the measured 
intensities of the strong reflections in reciprocal space, and to the expected
size of the object in real space. Subject to these constraints, 
it allows a thorough exploration of configuration space by iteratively modifying 
the phases of the weak reflections in reciprocal space and the signs of the small 
electron densities in real space. 
 
\section{Results for Noise-Free Simulations}

We have tested the effectiveness of this algorithm on a set of 630 simulated
diffraction patterns computed out to about 1 \AA \ resolution from random 
orientations of the small synthetic 
protein Chignolin, simulated from the atomic elements and coordinate data taken from 
the Protein Data Bank, and atomic scattering factors calculated from the 
relevant Cromer-Mann coefficients (Cromer and Mann, 1968). 

We then employed our common-line method to determine the Euler angles specifying
the relative orientations of each of the simulated diffraction patterns. A typical
comparison of the recovered angles with the known angles from the simulations 
is shown in Table \ref{table4}.

\begin{table}
\begin{center}
\begin{tabular}{||c|c|c|c||}
\cline{1-4}
Pair of Planes & Recovered/Actual $\Phi$ & Recovered/Actual $\Theta$ 
& Recovered/Actual $\Psi$ \cr
\cline{1-4}
(0,1) & 64.0/64.3 & 148.2/144.6 & 48.0/48.3 \cr
\cline{1-4}
(0,2) & 16.0/20.8 & 18.7/20.6 & 2.0/177.6 \cr
\cline{1-4}
(0,3) & 48.0/48.7 & 79.3/81.8 & 90.0/90.4 \cr
\cline{1-4}
(0,4) & 144.0/140.5 & 40.8/43.5 & 118.0/122.4 \cr
\cline{1-4}
(0,5) & 16.0/13.9 & 14.2/16.0 & 108.0/110.3 \cr
\cline{1-4}
(0,6) &  & Not found & \cr
\cline{1-4}
(0,7) & 174.0/174.2 & 138.9/138.2 & 100.0/99.6 \cr
\cline{1-4}
(0,8) & 90.0/90.0 & 92.3/87.9 & 168.0/169.7 \cr
\cline{1-4}
(0,9) & & Not found & \cr
\cline{1-4}
(0,10) & & Not found & \cr
\cline{1-4}
Mean error & 1.7 & 2.5 & 2.0 \cr
\cline{1-4}
\end{tabular}
\end{center}
\caption{Relative orientations of copies of a single molecule, as specified by a
set of Euler angles $\Phi$, $\Theta$, and $\Psi$, and the same angles recovered by
the identified common lines between pairs of the diffraction patterns (reciprocal 
space planes) labeled 0 to 10, and analytical formulae described in the text. Also
shown are the mean absolute errors in the determinations of these angles (all 
angles specified in degrees).}
\label{table4}
\end{table}

Occasionally the common-line search (section 2) does not succeed in accurately finding the 
Euler angles $\Phi$ and $\Psi$ relating a pair of diffraction patterns. If an angle 
$\Theta$ is calculated from such inexact values, the calculated argument of the arccosine in 
(\ref{Theta_ij}) may not lie in the range +1 to -1, and thus may not yield a value for 
the Euler angle $\Theta$. As shown by Table \ref{table4}, this is the case for 
the $\Theta$ angle relating diffraction patterns (0,9) and (0,10). In such cases, we simply
ignore the data in diffraction patterns 9 and 10. Proceeding this way, we were able
to determine self-consistent solutions for the orientations of 401 out of the 630
diffraction patterns simulated. For our sample of 11 diffraction patterns,
the mean accuracy of the Euler angle determination is about 2$^{o}$. 
We used the data of the 401 correctly oriented diffraction patterns to assign
intensities to an irregularly-spaced set of points in reciprocal space.

Is this sufficiently accurate? In order to answer this question one has to ask how
accurately one needs to determine these angles to correctly assign intensities in
each of the points of an oversampled 3D reciprocal-space grid. The required angular 
accuracy is thus determined by the angular extent of a reciprocal-space voxel of the 
highest resolution subtended at the origin of reciprocal space. Since the width of a 
reciprocal-space voxel is $1/(2L)$ (Huldt, {\it et al.}, 2003), where $L$ is a linear 
dimension of the molecule investigated, the angular resolution required is 
$1/(2L)/(1/R)=R/(2L)$ radians, where $R$ is 
the required resolution. For the example of the small protein modeled here, 
taking $L$=15 \AA \  and $R$=1 \AA \ , we may deduce that the required angular resolution 
is about 1/30$^{th}$ of a radian, or about 2$^{o}$. Table \ref{table4} shows that this is 
achieved.
 
\begin{figure}
\epsfxsize=3.0in
\epsffile{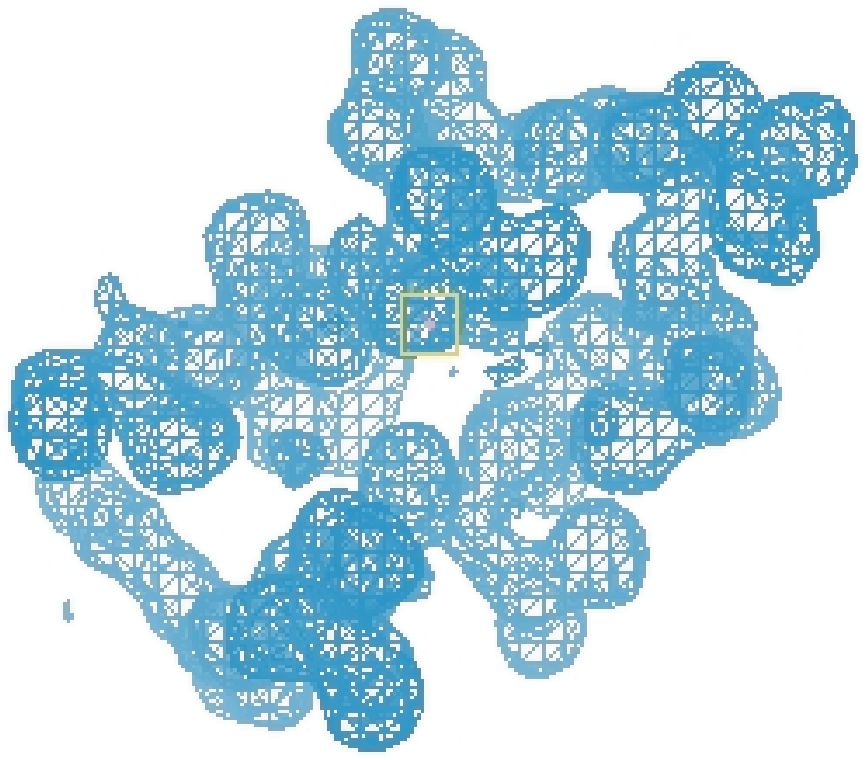}
\epsfxsize=2.9in
\epsffile{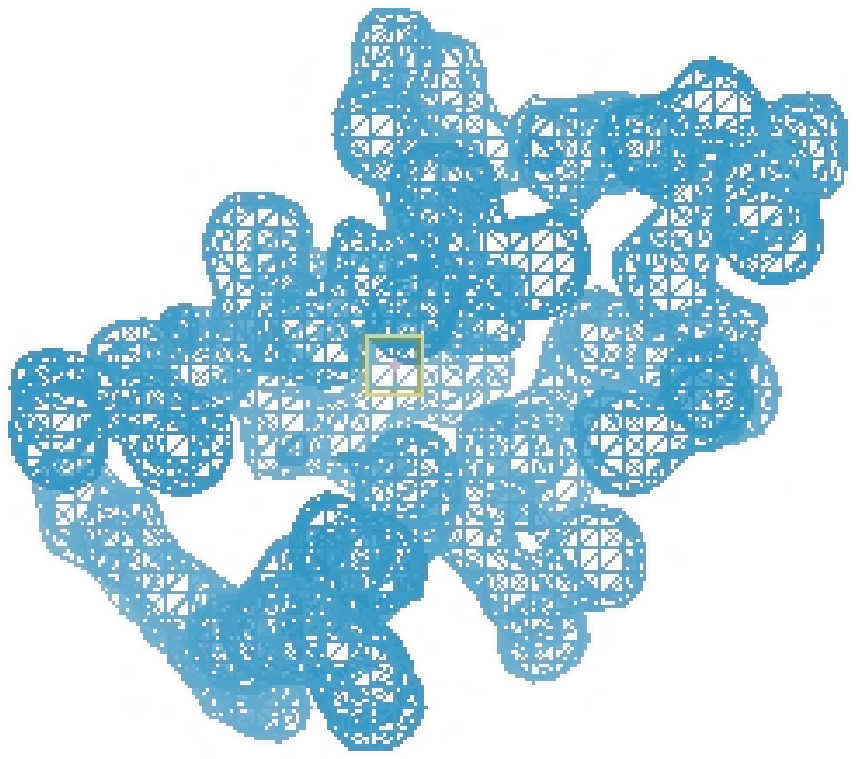}
\caption{{\it Electron density of protein Chignolin (PDB Entry: 1UAO) to about 1 
\AA \ resolution. Upper panel from PDB model. Lower panel
from from multiple diffraction patterns of molecule in random orientations.
The secondary structure is clearly visible.}}
\label{1uao}
\end{figure}

Application of the gridding algorithm of section 3 produced a set of diffraction 
intensities on such a uniform grid of points in 3D reciprocal space. 
Subsequent application of the iterative phasing algorithm of section 4 recovered 
the electron density distribution in the lower panel of 
Fig. \ref{1uao} in about 65 iterations of the phasing algorithm.

For purposes of comparison,
we also simulated the complex structure factors (amplitude and phase) on the
same oversampled 3D grid of reciprocal-space points as used in the iterative
phasing algorithm. An inverse Fourier transform of these (correct)
complex structure factors recovered the protein electron density distribution 
in the upper panel of Fig. \ref{1uao} at a resolution consistent with the extent 
of the diffraction data. 

The recovered electron density is in reasonable agreement with that of the 
starting model, with a correlation coefficient of 0.7 between the two electron
density distributions. 

\section{Effect of Shot Noise in Measured Diffraction Patterns}

Even with radiation from an ultra-bright source such as an XFEL, the expected number 
of detected photons per pixel of a diffraction pattern from a single biomolecule
is expected to be very small. Therefore, it is important to investigate the 
robustness of any algorithm to shot noise. We do this by assuming different mean 
photon counts per pixel $\bar{I_{0}}$ in the high resolution (or high-q) part
of the diffraction pattern. If $I_{0}$ is the expectation value of the photon 
count at any particular 
pixel, the actual number $I$ of detected photons is 
determined by the Poisson distribution 
\begin{equation}
p(I/I_{0})=\frac{I_{0}^I}{I!}e^{-I_{0}}
\end{equation}
where $p(I/I_{0})$ is the probability of measuring $I$ photons.
By comparing with the noise-free simulations, we investigated 
the effectiveness of the common-line algorithm in determining the relative Euler angles 
of the same diffraction patterns 0-10 of Table \ref{table4} for mean photon counts per pixel 
$\bar{I_{0}}=100$ and $\bar{I_{0}}=10$. 

The results of Table \ref{table4} were almost perfectly reproduced for
$\bar{I_{0}}=100$, but there was substantial deterioration of the fidelity of
the determined Euler angles for a mean photon count of $\bar{I_{0}}=10$,
(Table \ref{table5}).
 
\begin{table}
\begin{center}
\begin{tabular}{||c|c|c|c||}
\cline{1-4}
Pair of Planes & Recovered/Actual $\Phi$ & Recovered/Actual $\Theta$ 
& Recovered/Actual $\Psi$ \cr
\cline{1-4}
(0,1) & 68.0/64.3 & 150.6/144.6 & 52.0/48.3 \cr
\cline{1-4}
(0,2) & 14.0/20.8 & 18.1/20.6 & 4.0/177.6 \cr
\cline{1-4}
(0,3) & 48.0/48.7 & 85.6/81.8 & 92.0/90.4 \cr
\cline{1-4}
(0,4) & & Not found & \cr
\cline{1-4}
(0,5) & & Not found & \cr
\cline{1-4}
(0,6) & & Not found & \cr
\cline{1-4}
(0,7) & 174.0/174.2 & 144.3/138.2 & 100.0/99.6 \cr
\cline{1-4}
(0,8) & & Not found & \cr
\cline{1-4}
(0,9) & & Not found & \cr
\cline{1-4}
(0,10) & & Not found & \cr
\cline{1-4}
Mean error & 2.8 & 4.6 & 3.0 \cr
\cline{1-4}
\end{tabular}
\end{center}
\caption{Comparison of the determination of the relative Euler angles of the
same 11 diffraction patterns as for the noise-free case of Table \ref{table4}
for noisy diffraction patterns with mean photon count of 10 photons/pixel, with 
the (shot) noise modeled by a Poisson distribution}
\label{table5}
\end{table}

In the same subset of 10 diffraction patterns, the algorithm was able to determine just 
4 sets of relative Euler angles out of 10, with a mean angular accuracy of about 3.5$^{o}$. 
We stated earlier that the required angular resolution is $R/(2L)=2^o$, for 1 \AA 
\ resolution. This may be relaxed to about 4$^o$, if 2 \AA \ resolution is accepted. 
However, the fact that the orientations of less than half the diffraction patterns 
could be determined suggests that a mean detected 
photon count/pulse/pixel of 10 is close to the practical lower limit for the direct 
use of a common-line approach. Of course, our current simulations were performed for a 
small protein, and we have not explicitly tested the dependence of this limit on protein size.
However, it is of interest to note that a similar limit of counts per pixel is typical
for cryo-electron microscopy of biological entities.

The significance of these results becomes apparent on comparing these values of 
$\bar{I_{0}}$ with the estimated 
values of the same quantity under the usual assumptions of the incident beam flux from
an XFEL for two different values of the focussed beam diameter D, as
shown in Table \ref{table6}. 

\begin{table}
\begin{center}
\begin{tabular}{||c|c|c|c|c|c|c|c||}
\cline{1-8}
E(keV) & $\lambda$(nm) & 
\multicolumn{2}{c|}{$\sigma_C$ (mm$^2$/str.$\times$10$^{-22}$)} &
D ($\mu$m) & 
W (photons & 
\multicolumn{2}{c||}{n (ph/pulse/pixel)} \cr 
\cline{3-4} \cline{7-8}
 & & Small q & Large q & & /mm$^2$/pulse) & Small q & Large q \cr
\cline{1-8}
12.4 & 0.1 & 
2.87 & 0.26 & 
1 & 2.6$\times$10$^{18}$ & 
50 & 1.4$\times$10$^{-4}$ \cr
\cline{5-8} 
 & & & & 0.1 & 2.6$\times$10$^{20}$ &
5000 & 1.4$\times$10$^{-2}$ \cr
\cline{1-8}
\end{tabular}
\end{center}
\caption{Expected counts of detected photons/pulse/pixel for both small-q and large-q 
scattering by a 500 kDa protein with an XFEL source. E represents the photon 
energy, $\lambda$ its wavelength, $\sigma_C$ the typical differential scattering 
cross-sections for a C atom for small/large q, D the assumed diameter of a focussed
beam incident on the sample, W the photon fluence, and n the estimated scattered
photon count per pulse per detector pixel.}
\label{table6}
\end{table}

In compiling this Table, we assumed that the molecule consists of 
$N_{atom}$ non-H atoms (for the present purpose modeled as C atoms). We also distinguished 
between small-q and large-q scattering (where q is the scattering-induced momentum change
of an incident photon) for the following reason. There is a large difference between the 
expected photon count for small-q and for high-q scattering. 
Put simply, all electrons in the sample scatter more or less {\it in phase} in 
the low-q regime, thereby giving rise to a scattered intensity proportional to 
$N^{2}$, where $N$ is the number of electrons in the sample ($\simeq ZN_{atom}$, 
with $Z$ the average atomic number, and $N_{atom}$ the number of atoms), while
in the high-q regime, the scattered intensity is proportional to $N$. 

Values for the differential scattering cross section of 12.4 keV X-rays by a C atom 
for small q and large q scattering were taken from the tables on elastic photon-atom scattering
posted at the web site of the Lawrence Livermore National Laboratory
({\it http://www-phys.llnl.gov/Research/scattering}). Taking the effective width of a 
pixel as $\Delta k = 2 \pi/(2L)$ (Huldt {\it et al.}, 2003), where $L$ is a linear 
dimension of the molecule, implies a reciprocal
space pixel area of $(\Delta k)^2 = 4\pi^2/(4L^2)$. The solid angle subtended at
the sample by each pixel is then 
\begin{equation}
\Omega = (4\pi^2/4L^2)/k^2 = \lambda^2/4L^2 \mbox{ str.}
\label{omega}
\end{equation}
where $k$ is the wavenumber of the radiation, and $\lambda$ the wavelength. 
If $a$ is the average spacing of non-H atoms, we may take $(L/a)^3 \sim N_{atom}$, or
$L \sim aN_{atom}^{1/3}$. Substituting this value for $L$ in (\ref{omega}), we deduce
\begin{equation}
\Omega \sim \frac{\lambda^2}{4a^2} N_{atom}^{-2/3} \mbox{ str.}
\end{equation} 
For high q, the measured photon count per pulse per pixel ($n$) is
estimated as
\begin{equation}
n \sim \Omega N_{atom} W \sigma_C = \frac{\lambda^2}{4a^2} N_{atom}^{1/3} W \sigma_C,
\label{en}
\end{equation} 
with a similar expression for low q, but with a $N_{atom}^{4/3}$ dependence. Taking 
$N_{atom}=35,000$ (corresponding to a protein of approximately 500 kDa molecular weight), 
values for $\lambda$, $\sigma_C$, $D$, and $W$ given in Table \ref{table6}, and $a$ taken 
as 2 \AA, we deduce the values for $n$ for small/large momentum transfer q shown in the 
right-hand columns of Table \ref{table6}.

It is important to note that, even for a focussed beam diameter of 0.1 $\mu$m, 
the expected photon count per pixel for large-q data (needed for high-resolution structure 
determination) is approximately 3 orders of 
lower than the level at which the common-line method 
is able to reliably find the relative orientations of the diffraction patterns.
 
The estimates of Table \ref{table6} suggest that the photon counts in the low-q
region of a single diffraction pattern of a large protein
may be high enough to render the effects of shot noise negligible. However, it is 
unlikely that structural information directly available from low-q data will yield
anything more than the overall shape of the scattering object, as in the technique of
small angle X-ray scattering (SAXS). It is an open question whether a coarse 
orientating of patterns, which may be performed with the low-q data, will help
to orientate entire diffraction patterns sufficiently accurately to exploit 
the high-q data for high-resolution structure determination.  
  
\section{Discussion}

The ability to record and sort 2D diffraction patterns from individual molecules 
is important for a number of reasons. First and most obvious is the elimination
of the need for crystals. Second, and in our view equally important, is the 
potential to sort and 
separate diffraction patterns from different molecules or different molecular
conformations in the beam prior to structure recovery. The 
fact, for example, that 2D diffraction patterns from different molecules do not 
have common lines might allow the diffraction patterns to be separated into 
sets before further analysis, with each set designating a different type of 
molecule or molecular conformation.

This paper has been concerned with developing an algorithm to determine the structure 
of a single scattering entity (such as a protein, or nanoparticle) from multiple diffraction
patterns due to scattering from unknown random orientations of identical copies of the object. 
We have shown that an adaptation of a ``common-line'' algorithm from 3D electron 
microscopy/tomography is able to accomplish this task for noise-free diffraction patterns 
in the flat Ewald sphere limit. There is little doubt that an extension of such
a method to curved common-lines will similarly enable structure determination from
low-noise diffraction patterns at $\sim$ 1 \AA \ wavelengths characteristic of currently
planned XFELs (Hajdu {\it et al.}, 2000; Abela {\it et al.}, 2007).   

Of much greater concern is that fact that, even with the most powerful XFELs
currently envisaged, the expected number of scattered photons per high-q pixel
of a molecular diffraction pattern from a single radiation pulse is far too low for
the alignment approaches proposed so far. The common-line method relies on 
identifying similar intensity distributions along {\it single lines} in two low-intensity
(and thus high-noise) diffraction patterns. As such, it is hardly surprising that
it is very sensitive to noise. Our conclusion is that such a method requires
a mean photon count of at least 10 per pixel in the high-q region of 
a diffraction pattern, about 3 orders of magnitude greater than expected from
a proposed experiment with an XFEL (Table \ref{table6}).

We note that in the proposed experiments, the minimum photon
count per diffraction pattern orientation is not determined by the minimum required 
to reconstruct a satisfactory 3D image of the object from projections of known
orientations, as in conventional tomography, but rather by the need for correct 
classification and assembly of a 3D diffraction
volume from data in diffraction patterns alone. The minimum photon
count in the former case may be quite low, since the dose fractionation theorem
for 3D electron microscopy/tomography (Hegerl \& Hoppe, 1976; McEwen, Downing, 
\& Glaeser, 1985) states that ``A three-dimensional reconstruction requires the same 
integral dose as a conventional two-dimensional micrograph provided that the level of 
significance and resolution are identical''. This suggests that if there are M 
projections (or in our case, diffraction patterns) the photon count per pixel required
for an equally successful 3D reconstruction will be just 1/M of that for a single
projected image (in our case a single diffraction pattern). In the absence of
orientational information, this theorem does not help, because a much higher photon count 
(about 10 photons/pixel per orientational class) is needed for the successful assembly of 
a 3D diffraction volume suitable for structure solution. In short, the minimum photon count 
for correct classification and orientation is much higher than that needed for structure 
recovery of a single biomolecule. 

We now consider the possibility of classifying measured diffraction patterns into sets of 
similar orientations, and averaging their intensities to improve their signal-to-noise 
ratios. Bortel and Faigel (2007) find that successful classification of measured 
diffraction patterns of a protein modeled by 35,000 C atoms requires an incident photon 
fluence of 10$^{28}$ m$^{-2}$/pulse = 10$^{22}$ mm$^{-2}$/pulse. Comparison with 
our Table 
\ref{table6} shows that this is 100$\times$ the fluence expected from an XFEL beam focused 
down to a 0.1 $\mu$m diameter spot. Even if a more efficient method of classification were 
found, Table \ref{table6} indicates that the number of photons expected per
high-resolution pixel of the diffraction pattern is $\sim$ 10$^{-2}$, 
approximately 3 orders of magnitude smaller than that needed for the common-line method. 
This suggests the need for the summation of the data from about 1000 diffraction patterns 
per orientational class. Since Bortel and Faigel also find that at about 
10$^6$ classes are required for faithful recovery of the structure of such 
a molecule (assumed to be 100 \AA \ in diameter) to 3 \AA \ resolution, it will be 
necessary to measure $\sim$ 10$^{9}$ diffraction patterns. Assuming photon 
pulse and read-out rates of 100 Hz, this would require $\sim$ 10$^7$ seconds, 
or several months of continuous beam time for a single experiment.

We note that the classification problem is not eased for an even larger scattering object, 
such as a virus or
nanoparticle. Eq. (\ref{en}) suggests that the number of photons per detector pixel 
varies as $N_{atom}^{1/3}$. Thus, a scattering entity modeled by $N_{atom}$=10$^6$ C 
atoms, would scatter 3 times as many photons into each pixel. 
However, Bortel and Faigel (2007) estimate the number of required orientational classes for 
the structure solution of such an entity (assumed to be of 300 \AA \ diameter) to
3 \AA \ resolution is 2$\times$10$^7$, an extra factor of about 30 or so over the
500 kDa protein. Thus the
total number of diffraction patterns required, and hence time for data collection, would
be expected to increase to 10$^8$ seconds, i.e. several years for a single experiment.

Assembling a 3D intensity distribution from
the low-intensity diffraction patterns from single molecules obtainable from
single pulses of an XFEL may not be practical with the common-line method, the only 
approach mentioned in the literature so far (see e.g. the Technical Design Report of the 
European XFEL, Abela {\it et al.}, 2007). This calls for the development of entirely new
algorithms that perform structure solution by simultaneously acting on all the data of all 
measured diffraction patterns. Two such approaches have been suggested by the present 
authors 
(Ourmazd {\it et al.}, 2007; Saldin {\it et al.}, 2007) and will be the subject of 
forthcoming publications
\footnote{An alternative approach has been proposed by Spence {\it et al.} (2005) 
for improving the signal-to-noise ratio, in which the 
orientational alignment of the molecules is controlled by means of crossed laser beams.}.  

\section{Conclusions}

We have presented the first demonstration of an integrated algorithm to 
determine the electron density of a particle or large biomolecule, such as a protein, 
from a collection of 2D diffraction patterns, each from a molecule in an unknown random
orientation, as expected from the proposed X-ray scattering experiments with XFEL sources.
The method involves
first determining the relative orientations of the different 2D diffraction patterns,
interpolating the data onto a regular 3D 
Cartesian grid in reciprocal space at a sampling rate higher than the Nyquist frequency 
for the size of the molecule, determining the phases
associated with the measured amplitudes, and hence deducing the 3D electron
density of the molecule or nanoparticle. 
There are significant differences with similar algorithms developed previously for
3D electron microscopy, due to the absence of direct
phase information, and the ambiguities due to Friedel's Law. 
We have shown how these difficulties may be overcome, even in the limit
of a flat Ewald sphere, by the imposition of appropriate consistency conditions. 
These enable the determination of the relative orientations of the 
diffraction patterns, and hence the molecular structure, to within the usual
enantiomeric uncertainty. 

We have tested the algorithm with a computer simulation for a model protein,
at an X-ray wavelength short enough to justify the flat Ewald sphere approximation,
with and without Poisson noise for the detected photons. 
Adaptation of this algorithm to take account of curved common lines can readily 
extend the applicability of this approach to longer X-ray wavelengths.

Our simulations have highlighted an important limitation of a common-line
method for finding the relative orientations of diffraction patterns from
random orientations of a sample. Such methods depend on
comparing the intensity distributions along particular lines in two diffraction
patterns, thus using only a very small fraction of the available
data for each orientation determination. They are consequently very sensitive to noise. 

We find that the common-line method ceases to work reliably for mean photon counts
per pixel below about 10 in the high-q part of a diffraction pattern. These regions
contain the high-resolution information needed to resolve the secondary
structure of a protein. Since the scattering by a typical (500 kDa) protein of a pulse 
from a planned XFEL beam focused to a spot of 0.1 $\mu$m diameter is expected to produce 
some 1000$\times$ fewer photons per detector pixel, the 
use of a common-line method would seem to necessitate the classification and
averaging of at least 1000 low-intensity diffraction patterns per
orientational class to correctly assemble the scattered intensity distribution in 
3D reciprocal space. 

The method of classifying diffraction patterns into orientational classes
examined by Bortel and Faigel (2007) requires at least 100$\times$ the anticipated XFEL 
fluence. Even if superior classification methods were devised, the determination of 
the structure of a 100 \AA-wide molecule to 3 \AA \ resolution would require about 
10$^6$ orientational classes (Bortel and Faigel, 2007). Assuming pulse
and read-out rates of 100Hz, data collection for a 500 kDa protein would require 
several months of continuous beam time. 

%-------------------------------------------------------------------------
       % The back matter of the paper - acknowledgements and references
 
%-------------------------------------------------------------------------

       % Acknowledgements come after the appendices

We thank Veit Elser, Leonard Feldman, Paul Fuoss, John Spence, and Brian Stephenson
for helpful discussions. 

       % References are at the end of the document, between \begin{references}
       % and \end{references} tags. Each reference is in a \reference entry.
%\newpage
\vspace{0.3in}
{\Large {\bf References}}
\\ \\
Abela, R. {\it et al.} (2007). \emph{The European X-Ray Free-Electron
Laser: Technical Design Report,
http://xfel.desy.de/tdr/index\_eng.html}, edited by Altarelli, M. {\it et al.}, 
401-420.
\\ \\
Bortel, G. \& Faigel, G. (2007). \emph{J. Struct. Biol.}
\textbf{158}, 10-18.
\\ \\
Chapman, H. N., Barty, A., Marchesini, S., Noy, A., Hau-Riege, S. P.,
Cui, C., Howells, M. R., Rosen, R., He, H., Spence, J. C. H., Weierstall, U.,
Beetz, T., Jacobsen, C., \& Shapiro, D. (2006). \emph{J. Opt. Soc. Am.}
\textbf{23}, 1179-1200.
\\ \\
Chapman, H. N. (2007), private communication.
\\ \\
Cooley, J. W. \& Tukey, J. W. (1965). \emph{Math. of Comp.}
\textbf{19}, 297-301. 
\\ \\
Cramer, P., Bushnell, D. A., \& Kornberg, R. D. (2001). 
\emph{Science} \textbf{292}, 1863-1876.
\\ \\
Cromer, D. T. \& Mann, J. B. (1968). \emph{Acta Cryst.} A\textbf{24},
321-324.
\\ \\
Delaunay, B. (1934). \emph{Isvetia Academii Nauk SSSR, Otdelenie 
Mathematicheskikh Estestvennykh i Nauk} \textbf{7}, 793-800.
\\ \\
Elser, V. (2003). \emph{J. Opt. Soc. Am. A} \textbf{20}, 40-55. 
\\ \\
Gnatt, A. L., Cramer, P., Fu, J., Bushnell, D. A., \& Kornberg, R. D. (2001). 
\emph{Science} \textbf{292}, 1876-1882.
\\ \\
Farrow, N. A. \& Ottensmeyer, F. P. (1992). \emph{J. Opt. Soc. Am. A}
\textbf{9}, 1749-1760. 
\\ \\
Fenn, J. B. (2002). \emph{J. Biomolecular Techniques} \textbf{13}, 101-118.
\\ \\
Fienup, J. R. (1978), \emph{Opt. Lett.} \textbf{3}, 27-29.
\\ \\
Fienup, J. R. (1982), \emph{Appl. Opt.} \textbf{21}, 2758-2769.
\\ \\
Frank, J. (2006). \emph{Three-Dimensional Electron Microscopy of 
Macromolecular Assemblies}. Oxford University Press.
\\ \\
Fung, R., Shneerson, V. L., Lyman, P. F., Parihar, S. S.,
Johnson-Steigelman, H. T., \& Saldin, D. K. (2007). \emph{Acta Cryst.} A\textbf{63},
239-250.
\\ \\
Goncharov, A. B., Vainshtein, B. K., Ryskin, A. I., \& Vagin, A. A.  (1987).  
\emph{Sov. Phys. Crystallogr.} \textbf{32}, 504-509.
\\ \\
Hajdu, J., Hodgson, K., Miao, J., van der Spoel, D., Neutze, R., 
Robinson, C. V., Faigel, G., Jacobsen, C., Kirz, J., Sayre, D., Weckert, E.,
Materlik, G., \& Sz\"{o}ke, A. (2000). 
\emph{Structural Studies of Single Particles and Biomolecules, in LCLS: 
The First Experiments,
http://www-ssrl.slac.stanford.edu/lcls/papers/lcls\_experiments\_2.pdf}, 35-62.
\\ \\
Hegerl, R. \& Hoppe, W. (1976).
\emph{Z. Naturforschung} \textbf{31a}, 1717-1721.
\\ \\
Huldt, G., Sz\"{o}ke, A., \& Hajdu, J. (2003). 
\emph{J. Struct. Biol.} \textbf{144}, 218-227.
\\ \\
Jurek, Z., Faigel, G. \& Tegze, M. (2004). \emph{Eur. Phys. J.}
\textbf{29}, 217-229.
\\ \\
Kumpf, C., Marks, L. D., Ellis, D., Smilgies, D., Landemark, E.,
Nielsen, M., Feidenhans'l, R., Zegenhagen, J., Bunk, O., Zeysing, J. H.,
Su, Y. \& Johnson, R. L. (2001). \emph{Phys. Rev. Lett.} \textbf{86}, 
3586-3589.
\\ \\
Marchesini, S., He, H., Chapman, H. N., Hau-Riege, S. P., Noy, A., 
Howells, M. R., Weierstall, U. \& Spence, J. C. H. (2003). 
\emph{Phys. Rev. B} \textbf{68}, 140101(R).
\\ \\
McEwen, B. F., Downing, K. H., \& Glaeser, R. M. (1995).
\emph{Ultramicroscopy} \textbf{60}, 357-373.
\\ \\
Miao, J., Hodgson, K., \& Sayre D. (2001). \emph{Proc. Nat. Acad. Sci. 
U.S.A.}  \textbf{98}, 6641-6645.
\\ \\
Millane, R. P. (2003). \emph{Proc. Oceans 2003, CD ROM, IEEE} 2714-2719.
\\ \\
Neutze, R., Wouts, R., van der Spoel, D., Weckert, E., \& Hajdu, J. 
(2000).  \emph{Nature} \textbf{406}, 752-757.
\\ \\
Normille, D. (2006). \emph{Science} \textbf{314}. 751-752.
\\ \\
Oszl\'{a}nyi, G. \& S\"{u}to, A. (2003). \emph{Acta Cryst.} A\textbf{60},
134-141.
\\ \\
Oszl\'{a}nyi, G. \& S\"{u}to, A. (2004). \emph{Acta Cryst.} A\textbf{61},
147-152.
\\ \\
Ourmazd, A., Fung, R., Shneerson, V. L., and Saldin, D. K. (2007). 
\emph{Coherence 2007}, International Workshop on Phase Retrieval and Coherent
Scattering, Asilomar, California.
\\ \\
Saldin, D. K., Shneerson, V. L., Fung, R., Ourmazd, A. (2007). 
\emph{Coherence 2007}, International Workshop on Phase Retrieval and Coherent
Scattering, Asilomar, California.
\\ \\
Shapiro, D., Thibault, P., Beetz, T., Elser, V., Howells, M.,
Jacobsen, C., Kirz, J., Lima, E., Miao, H., Nieman, A. M., \& Sayre, D. (2005).
\emph{Proc. Nat.Acad. Sci. U.S.A.} \textbf{102}, 15343-15346.
\\ \\
Solem, J. C. \& Baldwin, G. C. (1982). \emph{Science}
\textbf{218}, 229-235.
\\ \\
Spence, J. C. H., Schmidt, K., Wu, J.S., Hembree, G., Weierstall, U., 
Doak, B., \& Fromme, P. (2005). \emph{Acta Cryst.} A\textbf{61}, 237-245.
\\ \\
Thibault, P., Elser, V., Jacobsen, C., Shapiro, D., \& Sayre, D.
(2006). \emph{Acta Cryst.} A\textbf{62}, 248-261.
\\ \\ 
van Heel, M. (1987). \emph{Ultramicroscopy} \textbf{21}, 111-124.
\\ \\
Williams, G. J., Pfeifer, M. A., Vartanyants, I. A., \& 
Robinson, I. K. (2003). \emph{Phys. Rev. Lett.} \textbf{90}, 175501.
\\ \\ 
Wu, J. S., Spence, J. C. H., O'Keeffe, M. A., \& Groy, T. L. (2004a). 
\emph{Acta Cryst.} A\textbf{60}, 326-330.
\\ \\
Wu, J. S., Weierstall, U., Spence, J. C. H., \& Koch, C. T. (2004b).
\emph{Opt. Lett.} \textbf{29}, 2737-2739.
\\ \\
Wu, J., Weierstall, U., \& Spence, J. C. H. (2005). 
\emph{Nature Materials} \textbf{4}, 912-916. 

%-------------------------------------------------------------------------
       % TABLES AND FIGURES SHOULD BE INSERTED AFTER THE MAIN BODY OF THE TEXT
 
%-------------------------------------------------------------------------

       % Simple tables should use the tabular environment according to this
       % model

%\begin{table}
%\caption{Caption to table}
%\begin{tabular}{llcr}      
% Alignment for each cell: l=left, c=center, r=right
% HEADING    & FOR        & EACH       & COLUMN     \\
%\hline
% entry      & entry      & entry      & entry      \\
% entry      & entry      & entry      & entry      \\
% entry      & entry      & entry      & entry      \\
%\end{tabular}
%\end{table}

       % Postscript figures can be included with multiple figure blocks

%\begin{figure}
%\caption{Caption describing figure.}
%\includegraphics{fig1.ps}
%\end{figure}

\end{document}